\documentclass[fleqn,usenatbib]{mnras}
\usepackage{ae,aecompl}
\usepackage{graphicx}	
\usepackage{amsmath}	
\usepackage{amssymb}	
\usepackage{color}

\title[Cosmic ray driven Galactic winds]{Cosmic ray driven Galactic winds}

\author[]{
S. Recchia $^{1}$\thanks{E-mail: sarah.recchia@gssi.infn.it}
P. Blasi,$^{2,1}$\thanks{E-mail: blasi@arcetri.astro.it}
G. Morlino $^{1,2}$\thanks{E-mail: giovanni.morlino@gssi.infn.it}
\\
$^{1}$ Gran Sasso Science Institute (INFN), Viale F. Crispi 7 - 67100 L' Aquila, Italy\\
$^{2}$ INAF/Osservatorio Astrofisico di Arcetri, Largo E. Fermi, 5 - 50125 Firenze, Italy
}

\date{Accepted XXX. Received YYY; in original form ZZZ}

\pubyear{2016}

\begin{document}
\label{firstpage}
\pagerange{\pageref{firstpage}--\pageref{lastpage}}
\maketitle

\begin{abstract}
The escape of cosmic rays from the Galaxy leads to a gradient in the cosmic ray pressure that acts as a force on the background plasma, in the direction opposite to the gravitational pull. If this force is large enough to win against gravity, a wind can be launched that removes gas from the Galaxy, thereby regulating several physical processes, including star formation. The dynamics of these cosmic ray driven winds is intrinsically non-linear in that the spectrum of cosmic rays determines the characteristics of the wind (velocity, pressure, magnetic field) and in turn the wind dynamics affects the cosmic ray spectrum. Moreover, the gradient of the cosmic ray distribution function causes excitation of Alfv\'en waves, that in turn determine the scattering properties of cosmic rays, namely their diffusive transport. These effects all feed into each other so that what we see at the Earth is the result of these non-linear effects. Here we investigate the launch and evolution of such winds, and we determine the implications for the spectrum of cosmic rays by solving together the hydrodynamical equations for the wind and the transport equation for cosmic rays under the action of self-generated diffusion and advection with the wind and the self-excited Alfv\'en waves. 
\end{abstract}
\begin{keywords}
keyword1 -- keyword2 -- keyword3
\end{keywords}



\section{Introduction}
\label{sec:intro}
The possibility of a galaxy to launch winds has attracted attention for many different reasons. For instance, star formation is regulated by the amount of gas available, and winds affect the availability of such gas. In fact, galactic models that do not include feedback processes suffer from overpredicting the amount of baryons and star formation rates \citep{Crain:2007MNRAS.377...41C,Stinson:2013MNRAS.428..129S}. 
Winds also pollute Galactic halos with hot dilute plasma that may provide an important contribution to the number of baryons in the Universe\citep{Kalberla:1998A&A...332L..61K,Kalberla:2008A&A...487..951K,Miller:2013ApJ...770..118M}. Such gas might in fact have already been detected \cite[]{1995ApJ...454..643S} in the form of a X-ray emitting plasma with temperature of several million degrees, and possibly associated with a Galactic wind by \cite{1994Natur.371..774B,1999A&A...347..650B}  \cite[see also][]{Everett:2008p3580}. Finally winds can affect the transport of cosmic rays (CRs) in a galaxy, by advecting them away from their sources. 

Galactic winds may be thermally-driven, namely powered by core-collapse SNe \cite[see e.g.][]{Chevalier:1985Natur.317...44C} or momentum-driven, powered by starburst radiation \citep{Scoville:2003JKAS...36..167S,Murray:2005ApJ...618..569M}. These two mechanisms of wind launching are thought to be at work in starburst galaxies and galaxies with active nuclei \cite[see, e.g.,][for a comprehensive review]{Veilleux:2005ARA&A..43..769V}. On the other hand in a galaxy like the Milky Way, winds are unlikely to be due to such processes because thermal and radiation pressure  gradients are expected to be too small. A possible exception is the innermost part of the Galactic Center region where the recent discovered Fermi Bubbles may have their origin through direct bursts from Sgr A$^*$ \citep{Cheng:2011ApJ...731L..17C,Zubovas:2011MNRAS.415L..21Z} or because of past starburst activities \citep{Lacki:2014MNRAS.444L..39L}. On the other hand, CRs can play an important role in launching winds because of the gradient that their pressure develops as a consequence of the gradual escape of CRs from the Galaxy. The force $-\nabla P_{\rm CR}$ associated with such gradient is directed opposite to the gravitational force, and in certain conditions the plasma above and below the disc can be lifted off to form a CR driven wind. Notice that the gravitational force may be dominated by the dark matter component or the baryonic (gas and stars) components depending on the location. The force exerted by CRs depends in a complicated manner on the density of sources of CRs but also on non-linear processes of excitation of Alfv\'en waves through streaming instability. Both the force induced by CRs on the background plasma and the streaming instability induced by CRs depend on the gradient in CR density. In turn the distribution function of CRs is affected by their transport: diffusion is self-regulated through the production of Alfv\'en waves, and advection is determined by the velocity of the wind, if any is launched, and by the Alfv\'en waves' velocity, directed away from the sources of CRs. This complex interplay makes the problem non-linear and in fact the full problem of calculating at the same time the properties of the wind and the spectrum of CRs as a function of the location in the Galaxy has never been solved before. The first pioneering attempt to describe the hydrodynamics of a CR driven wind was described in a paper by \cite{Ipavich:1975p3566}, where the author used a spherically symmetric model of the Galaxy and considered only baryons and stars for the calculation of the gravitational potential. Later \cite{Breitschwerdt:1991A&A...245...79B} presented an extensive discussion of the hydrodynamics of CR driven winds: dark matter was included and a realistic geometry of the wind was considered, in which the launch takes place at some distance from the Galactic disc and proceeds in a roughly cylindrical symmetry out to a distance of about $\sim 15$ kpc, where the flow opens up into a spherical shape. Such a geometry became a milestone for future calculations of CR driven winds and is adopted also in our paper, as discussed below. The calculations of \cite{Breitschwerdt:1991A&A...245...79B} treated CRs as a fluid, hence no information on the spectrum of CRs was retained. The important role of wave damping in the wind region was also discussed by \cite{Breitschwerdt:1991A&A...245...79B}, although only in the simplified case of a spherical outflow. As mentioned above, \cite{Breitschwerdt:1991A&A...245...79B} assumed that the wind is launched some distance away from the disc of the Galaxy and this assumption raises the issue of what happens in the region between the disc and the base of the wind, a problem of both mathematical and physical importance, that was discussed by \cite{Breitschwerdt:1993p3640} and that will be central to our paper as well. 

The paper by \cite{Breitschwerdt:1991A&A...245...79B} represented a milestone in the investigation of CR driven winds and their calculation was adopted in much of the future literature in the field, including the recent work by \cite{Everett:2008p3580} that used the model of a CR driven wind (including damping) to explain the observed Galactic soft X-ray emission. 

The dynamical role of CRs in launching the winds was also studied via purely hydrodynamical simulations \citep{Uhlig:2012MNRAS.423.2374U,Booth:2013ApJ...777L..16B,Salem:2014MNRAS.437.3312S} and through MHD simulations \citep{Girichidis:2016ApJ...816L..19G,Peters:2015ApJ...813L..27P,Ruszkowski2016}. These simulations, with their progressive level of sophistication, demonstrated that CRs play an important role in wind launching. Nevertheless, all this bulk of work treated CRs as a fluid, thereby not providing any information on the CR spectrum.

The first calculation of the spectrum of CRs in a Galaxy with a CR driven wind was made by \cite{Ptuskin:1997A&A...321..434P}, where the rotation of the Galaxy over cosmological time scales was also taken into account \cite[]{Zirakashvili:1996A&A...311..113Z}. The authors used a simplified approach to the characteristics of the wind (for instance the assumption that the advection velocity for small heights above the disc scales linearly with distance from the disc) to infer some general implications for the spectrum of CRs. To our knowledge, the paper by \cite{Ptuskin:1997A&A...321..434P} represents the only attempt to account for both a CR driven wind and its effects of the transport of CRs. Much work has been done in later years on treating the wind in a more realistic way, including the recent time dependent approach to the wind evolution \citep{Dorfi:2012A&A...540A..77D} and numerical simulations of the wind dynamics \citep{Uhlig:2012MNRAS.423.2374U} and chemical evolution \cite[]{Girichidis:2016ApJ...816L..19G}, but none of these efforts included the kinetic description of CRs. 

In this paper we present the first semi-analytical calculation of the hydrodynamics of a CR driven wind, including the self-generation of Alfv\'en waves through streaming instability and their damping through non-linear Landau damping, and the calculation of the diffusive-convective transport of CRs in such wind. Throughout the paper, we assume that the general topology of the magnetic field lines is such that the transport of the wind and of CRs occurs along the field lines (parallel geometry).

At present, for the sake of simplicity, we decided to leave rotation out of this calculations, but its introduction is straightforward from the technical point of view. The spectrum of CRs at the position of the Earth is calculated for several cases of interest, so as to emphasise that several of the cases that are considered feasible from the point of view of hydrodynamics actually lead to spectra that are not consistent with observations. Special attention is devoted to the discussion of the role of the {\it near disc} region, where waves cannot be produced because of ion-neutral damping, and CR scattering, if any is present, is most likely guaranteed by some sort of turbulence directly injected by the same sources of CRs, for instance supernova explosions. 

We find that the spectrum of CRs observed at the Earth in the low energy regime ($E\lesssim 1$ TeV) is mainly affected by the wind launching and evolution, because of the interplay between advection with the wind (and Alfv\'en waves) and self-generated diffusion. In general, at $E\lesssim 10$ GeV advection is dominant, while a relatively steep spectrum is found at higher energies. For $E\gtrsim 1$ TeV the CR transport is dominated by what happens in the near-disc region. In general this situation may lead to a spectral change in the TeV range, that under some conditions, may resemble the hardening observed by PAMELA \cite[]{pamhard} and AMS-02 \cite[]{amshard}, at least qualitatively. 

The paper is organised as follows: in \S \ref{sec:method} we describe the basic formalism and the mathematical methods used to compute the structure of the wind in the presence of self-generated waves. The details of how to solve the hydrodynamical equations and the CR transport equation are provided in \S \ref{sec:Hydro} and \S \ref{sec:Kin} respectively. Our results are discussed in \S \ref{sec:results}, for our reference model (\S \ref{sec:reference}) and for an alternative scenario devised to illustrate the importance of the near-disc region (\S \ref{sec:kolmo}). We draw our conclusions in \S \ref{sec:discuss}. 

\section{A semi-analytical approach to CR-driven winds}
\label{sec:method}

The dynamics of winds in the presence of CRs is described by the equations of conservation of mass, momentum and energy for the wind itself and by the transport equation for CRs. The two are coupled to each other in three ways: 1) the gradient of the CR pressure acts as a force on the plasma; 2) Alfv\'en waves excited by CR streaming are quickly damped through non-linear Landau damping (NLLD), thereby resulting in heating of the plasma in the wind; 3) the diffusion term in the CR transport equation leads to an effective contribution to the enthalpy. In the assumption that damping occurs on time scales much shorter than any other process, definitely justified for NLLD, one can write the equations of hydrodynamics of the wind (see Appendix \ref{sec:appendix Hydro} for their derivation) as:
\begin{align}
&\vec{\nabla}\cdot (\rho \vec{u})=0, \label{eq:H-mass} \\
&\rho(\vec{u}\cdot\vec{\nabla})\vec{u}=-\vec{\nabla}(P_g+P_c)-\rho\vec{\nabla}\Phi, \label{eq:H-u}\\
&\vec{u}\cdot\vec{\nabla}P_g = \frac{\gamma_g P_g}{\rho}\vec{u}\cdot\vec{\nabla}\rho - (\gamma_g-1)\vec{v_A}\cdot\vec{\nabla}P_c, 
\label{eq:H-Pg}\\
&\vec{\nabla}\cdot\left[\rho \vec{u}\left(\frac{u^2}{2}+\frac{\gamma_g}{\gamma_g-1}\frac{P_g}{\rho} + \Phi\right)\right]=
-(\vec{u}+\vec{v}_A)\cdot\vec{\nabla}P_c, \label{eq:H-energy}\\
&\vec{\nabla}\cdot\left[(\vec{u}+\vec{v}_A)\frac{\gamma_c P_c}{\gamma_c -1} -\frac{\overline{D}\vec{\nabla}P_c}{\gamma_c-1}\right]=
(\vec{u}+\vec{v}_A)\cdot\vec{\nabla}P_c, \label{eq:H-Pc}\\
&\vec{\nabla}\cdot\vec{B}=0 \label{eq:H-B},
\end{align}
where $\rho(z)$, $\vec{u}(z)$ and $P_g(z)$ are the gas density, velocity and pressure and $\vec B$ is the magnetic field in the wind, while $P_{c}(z)$ is the CR pressure and $\gamma_{c}$ is the adiabatic index of the CR gas. Notice that $\gamma_{c}(z)$ is actually calculated locally (as a function of $z$) from the distribution function $f(p,z)$ that solves the CR transport equation. 

We introduced the Alfv\'en velocity $\vec{v}_A(z)$, while $\Phi(R_0, z)$ is the gravitational potential of the Galaxy (see Sec. \ref{subsec:grav}). 
The fact that the wave pressure does not appear in the equations reflects the assumption of fast damping, which results in small wave pressure compared to the gas and CR pressures.\\
The transport of CRs is described by the advection-diffusion equation:
\begin{equation}\label{eq:K-CR transport}
\vec{\nabla}\cdot\left[D\vec{\nabla}f\right]-(\vec{u}+\vec{v}_A)\cdot\vec{\nabla}f+\vec{\nabla}
\cdot(\vec{u}+\vec{v}_A)\frac{1}{3}\frac{\partial f}{\partial \ln p}+Q=0,
\end{equation}
where $f(\vec r,p)$ and $D(\vec r,p)$ are the CR distribution function and diffusion coefficient as functions of position $\vec r$ and momentum $p$. The term $Q$ represents injection of CRs in the Galaxy, that we assume to be limited to the Galactic disc. The average diffusion coefficient that appears in equation \ref{eq:H-Pc} is defined as:
\begin{equation}
\overline{D} (\vec r)= \frac{\int_{0}^{\infty} dp~ p^{2} T(p) D(\vec r,p) \nabla f}{\int_{0}^{\infty} dp ~p^{2} T(p) \nabla f},
\end{equation}
and $T(p)$ is the kinetic energy of particles with momentum $p$.

\begin{figure}
	\includegraphics[width=\columnwidth]{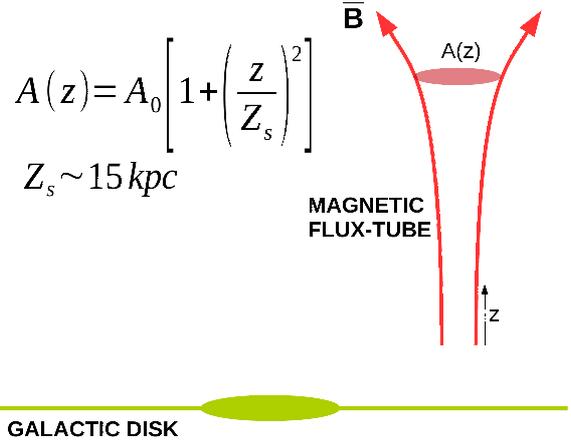}
    \caption{Flux-tube geometry for the magnetic field.}
    \label{fig:flux-tube}
\end{figure}

In the following we adopt the same geometry of the wind flow as initially introduced by \cite{Breitschwerdt:1991A&A...245...79B}, and used by many authors afterwards. The assumption is that the wind is launched from a surface at distance $z_{0}$ above (and below) the disc, and that it retains a roughly cylindrical geometry out to a distance $Z_{s}$ of the order of the radius of the disc. At larger distances the flow opens into a spherical shape. The surface of the wind is then assumed to be in the form:
\begin{equation}
A(z)=A_{0} \left[ 1+ \left( \frac{z}{Z_s}\right)^{2}\right],
\end{equation}
which is only function of one spatial coordinate, $z$. This makes the problem effectively one-dimensional as long as all quantities are assumed to depend only on $z$. This simplifies the conservation equations that can be easily shown to lead to the following expressions (see Appendix \ref{sec:appendix_tube}):
\begin{align}
& \rho u A=\textnormal{const}, \label{eq:Hcal-mass} \\
& AB= \textnormal{const},\label{eq:Hcal-B}\\
& \frac{du}{dz}=u\frac{c_{*}^2\frac{1}{A} \frac{dA}{dz} -\frac{d\Phi}{dz}}{u^2-c_{*}^2}, \label{eq:Hcal-wind eq}\\
& \frac{dP_g}{dz}=\gamma_g\frac{P_g}{\rho}\frac{d\rho}{dz}-(\gamma_g -1)\frac{v_A}{u}\frac{dP_c}{dz} \label{eq:Hcal-Pg}\\
&\frac{dP_c}{dz}=\frac{\gamma_c P_c}{\rho}\frac{2u+v_A}{2(u+v_A-\mathcal{D})}\frac{d\rho}{dz}, \label{eq:Hcal-Pc}\\
& c_{*}^2=\gamma_g \frac{P_g}{\rho} + \gamma_c \frac{P_c}{\rho} \left[ 1-(\gamma_g-1)\frac{v_A}{u} \right]\frac{2u+v_A}{2(u+v_A -\mathcal{D})}, \label{eq:Hcal-cs}\\
& \mathcal{D}=\frac{\frac{d}{dz}\left(A\overline{D}\frac{dP_c}{dz}\right)}{A\frac{dP_c}{dz}}. \label{eq:Hcal-DD}
\end{align}
One can easily recognise the generalised sound speed $c_{*}$ introduced by \cite{Breitschwerdt:1991A&A...245...79B}, where however two important differences appear: first, the non-adiabatic heating induced by wave damping, which has important implications for the wind launching. Second, the effective velocity term $\mathcal{D}$ which accounts for the finite diffusivity of CRs. In the calculations of \cite{Breitschwerdt:1991A&A...245...79B}, it was assumed that $\overline{D}=0$, although one can check a posteriori that in general this assumption is not justified, in that there are locations where such term is at least comparable with all others, so that neglecting it leads to an incorrect solution of the problem.

The transport equation of CRs also gets simplified with the flux geometry assumed above:
\begin{align}
\frac{\partial}{\partial z}& \left[ A(z)D(z,p)\frac{\partial f(z,p)}{\partial z}\right] -\left[A(z)U(z)\frac{\partial f(z,p)}{\partial z}\right] \label{eq: CR transport equation flux tube}\\ \nonumber
& +\frac{d[A(z)U(z)]}{dz}\frac{1}{3}\frac{\partial f(z,p)}{\partial \ln p}+A(z)Q(z,p)=0,
\end{align}
where we introduced the advection velocity $U(z)=u(z)+v_{A}(z)$. In general the diffusion coefficient $D(z,p)$ is assumed to be determined by the process of self-generation of Alfv\'en waves excited by CRs through streaming instability, although we will consider situations in which we relax this assumption. Since the self-generated perturbations in the magnetic field are relatively weak, one can still use quasi-linear theory to write the diffusion coefficient as:
\begin{equation}\label{eq:D self-gen}
D(z,p)=\left.\frac{1}{3}\frac{v(p)r_L(z,p)}{\mathcal{F}}\right|_{k_{\rm res}=1/r_L},
\end{equation}
where $\mathcal{F}$ is the normalized energy density per unit logarithmic wavenumber $k$, calculated at the resonant wavenumber $k_{\rm res}=1/r_L(p)$. The local value of $\mathcal{F}$ is determined by the balance between the CR-driven growth of Alfv\'en waves and their damping. In the region where the background gas is totally ionized, waves are damped through NLLD at a rate \cite[]{2003A&A...403....1P}:
\begin{equation}
\Gamma_{\rm D} = (2c_k)^{-3/2}kv_A \mathcal{F}^{1/2}. \label{eq:GAMMA NLD} 
\end{equation}
On the other hand the growth occurs at a rate that is given by \cite{Skilling:1971p2173}:
\begin{equation}
\Gamma_{\rm CR}=\frac{16\pi^2}{3}\frac{v_A}{\mathcal{F} B^2}\left[p^4v(p)\left| \frac{\partial f}{\partial z} \right| \right]_{p=p_{res}}. \label{eq:GAMMA CR}
\end{equation}
Equating $\Gamma_{\rm D}= \Gamma_{\rm CR}$ and using equation (\ref{eq:D self-gen}) one obtains:
\begin{equation}\label{eq:wave spectrum}
\mathcal{F}(z,p)=2c_k\left[\frac{p^4v(p) \left| \frac{\partial f}{\partial z} \right| 
\frac{16\pi^2}{3}r_L(z,p)}{B^2(z)} \right]^{2/3}.
\end{equation}

In \S \ref{sec:Hydro} and \S \ref{sec:Kin} we will describe the method adopted to solve the hydrodynamical equations and the transport equation respectively. Here we describe the way the two pieces are linked together in an iterative procedure. We initiate our computational procedure by solving the equation for the wind structure, by assuming that all quantities are known at the base of the wind, and by requiring that $\overline{D}=0$. This first step allows us to derive a guess for the launching velocity of the wind $u_{0}$ and the wind properties (velocity, pressure of the gas, CRs and waves). At this point this guess is used as an input for the quantities entering the transport equation (\ref{eq: CR transport equation flux tube}): most important, the wind velocity $u(z)$ and the Alfv\'en speed $v_{A}(z)=B(z)/\sqrt{4\pi \rho(z)}$ are calculated from the output of the hydrodynamical part of the calculation. The solution of equation (\ref{eq: CR transport equation flux tube}) leads to the spectrum of CRs as a function of location $z$ and momentum $p$, $f(z,p)$, and to knowing the diffusion coefficient $D(z,p)$ as due to self-generated Alfv\'en waves. Notice that in general the CR pressure obtained by integrating the distribution function $f(z,p)$ (and especially the CR pressure at the base of the wind) is different from the one assumed in order to determine the wind structure. This leads to the second iteration: the updated CR pressure at the base is used to compute the structure of the wind and the whole calculation is repeated. Moreover, the diffusion coefficient and the distribution function are used to compute $\gamma(z)$ and $\overline{D}$. Notice that the latter quantity is no longer bound to vanish, as assumed in previous calculations by \cite{Breitschwerdt:1991A&A...245...79B}. This iterative scheme continues until the spectrum of CRs reaches convergence at all distances and at all momenta. As a consequence the CR pressure at the base converges to the actual solution, so it does the whole structure of the CR induced wind. Clearly, if the goal is to apply this calculation to the case of our own Galaxy at the location of the Sun, $R_{\odot}$, then the CR injection is modified so as to reach a solution where the CR pressure is the same as observed. It is very important to stress that this condition alone is not sufficient to guarantee that the CR spectrum observed at the Earth is also consistent with observations: in fact, a generic wind solution leads to spectra of CRs at the Earth location that are quite different from the observed ones, mainly as a result of the strong advection with the wind, that corresponds to exceedingly hard spectra at low energy. We will discuss the qualitative comparison between the wind solutions and the observed CR spectra in \S \ref{sec:results}.

Since the properties of the wind depend in a rather sensible way upon the properties of the gravitational potential, in \S \ref{subsec:grav} we describe in detail our assumptions on the contribution of disc and bulge content and dark matter to the gravitational potential. 

\subsection{Galactic gravitational potential} 
\label{subsec:grav}

The gravitational potential of the bulge-disk component is modelled following \cite[]{Miyamoto:1975PASJ...27..533M}: 
\begin{equation}\label{eq:bulge-disk}
\Phi_{\rm B,D}(R_0, z)= - \sum_{i=1}^2 \frac{G M_i}{\sqrt{R_0^2 + \left(a_i + \sqrt{z^2 + b_i^2}\right)^2}},
\end{equation}
where $z$ and $R_0$ are the distance from the galactic disk and the galactocentric distance respectively. Following \cite[]{Breitschwerdt:1991A&A...245...79B}, we assumed for the bulge 
$(M_1, a_1, b_1)=$($2.05\times 10^{10}M_{\odot}$, 0.0 kpc, 0.495 kpc) and $(M_2, a_2, b_2)=$($2.547\times 10^{11}M_{\odot}$, 7.258 kpc, 0.520 kpc) for the disk.

Dark matter distribution is assumed to be well described by a Navarro-Frenk-White (NFW) profile \cite[]{nfw}:
\begin{equation}
\rho_{\rm DM} = \frac{\rho_{0}}{x(1+x)^{2}},
\end{equation}
where $x=r/r_{c}$ and $r_{c}$ is the radius of the core of the distribution. The two quantities $\rho_{0}$ and $r_{c}$ are calculated by requiring that the total mass of the halo is $M_{\rm vir}=10^{12}M_{\odot}$ \cite[]{Salucci:2013JCAP...07..016N} and the energy density of dark matter at the position of the Sun is $0.3$~GeV~cm$^{-3}$. The dark matter halo is assumed to extend out to a maximum distance that equals the viral radius $r_{\rm vir}\approx 300$ kpc. 

\begin{figure}
	\includegraphics[width=\columnwidth]{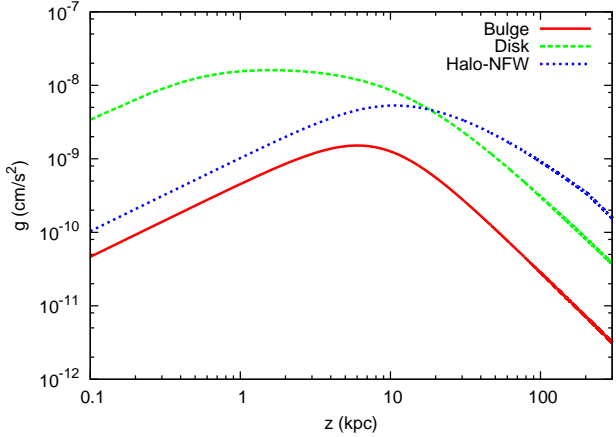}
    \caption{Gravitational acceleration as due to the three components: the Bulge, the Disk and the Dark Matter Halo. All curves are computed at the Sun's position ($R_{\odot} =8.5$ kpc).}
    \label{fig:GravPot}
\end{figure}

The Dark Matter halo potential corresponding to this spatial distribution reads: 
\begin{align}
\Phi(x)= & -4\pi G\rho_s r_s^2\left[ \frac{\ln(1+x)}{x} -\frac{\ln(1+x_{\rm vir})}{x_{\rm vir}}\right] -\label{eq:NFW} \\ \nonumber
         &- \frac{GM_{\rm vir}}{r_s}\frac{1}{x_{\rm vir}} \qquad\qquad\qquad\qquad (r \leq r_{\rm vir})  \\ \nonumber
         \\ \nonumber
\Phi(x)= & - \frac{GM_{\rm vir}}{r_s}\frac{1}{x}. \qquad\qquad\qquad\qquad (r > r_{\rm vir}) 
\end{align}
A plot of the acceleration associated with the different contributions to the Galactic gravitational potential at the position of the Sun $R_{\odot}=8.5$ kpc is shown in Fig. \ref{fig:GravPot} as a function of the height $z$ from the disc. One can see that at $R_{\odot}$ the potential is dominated by the disc component out to a distance $z\sim 10$ kpc, while at larger distances the dark matter contribution becomes more important.

\section{Identification of the wind solution}
\label{sec:Hydro}

Here we discuss the procedure adopted for the determination of the wind solution for a given set of conditions at the base of the wind. We recall that, as discussed above, this calculation is embedded in an iterative computation that is repeated until convergence of the overall structure of the wind and of the CR spectrum is achieved. 

We look for a wind solution of equations (\ref{eq:Hcal-mass})-(\ref{eq:Hcal-DD}), namely a solution for the flow velocity $u(z)$ that shows a transition from subsonic ($ u<c_{*}$) to supersonic ($u > c_{*}$) motion, where $c_{*}(z)$ is the compound sound speed as a function of $z$. 

\begin{figure}
	\includegraphics[width=\columnwidth]{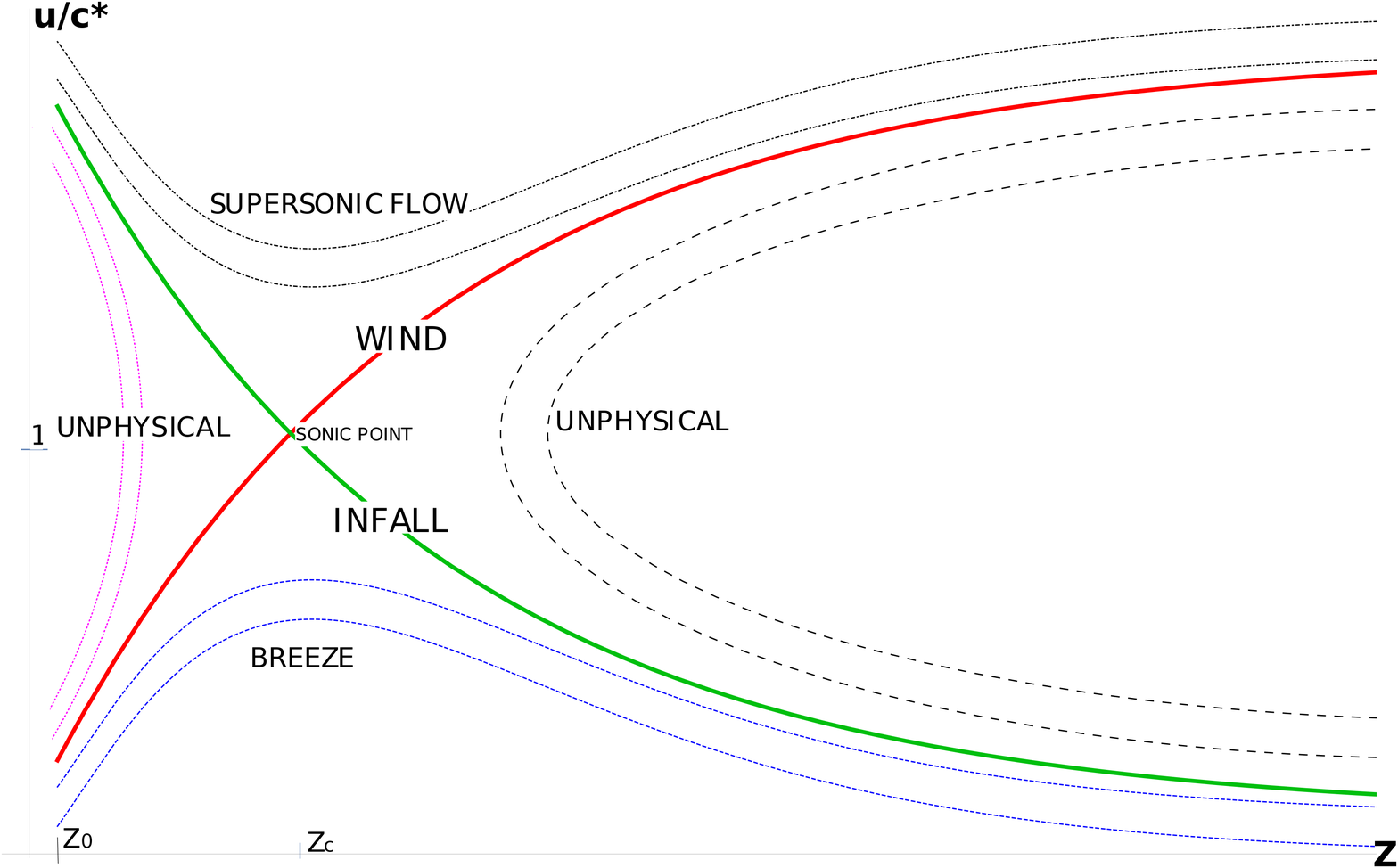}
    \caption{Topology of the solutions of the hydrodynamics equations: $z_0$ indicates the base of the wind, while $z_c$ is the critical point, where $u(z_{c})=c_{*}(z_{c})$. The (red) solid line represents a wind solution, that starts as subsonic and becomes supersonic at the critical point. The branches of the solutions that are unphysical are labelled as such.}
    \label{fig:topology}
\end{figure}

The topology of the solutions \cite[see also][]{Breitschwerdt:1991A&A...245...79B}) of equation (\ref{eq:Hcal-mass})-(\ref{eq:Hcal-DD}) depends on the nature of the critical points of the wind equation (\ref{eq:Hcal-wind eq}), namely the points in which the velocity derivative has zero numerator $(c_{*}^2= \frac{d\Phi}{dz}/\frac{1}{A} \frac{dA}{dz})$  and/or zero denominator ($u^2=c_{*}^2$), and is schematically represented in Fig. \ref{fig:topology}. The point for which both numerator and denominator are zero is the critical (sonic) point and it corresponds to the location where the flow velocity equals the compound sound speed, $u=c_{*}$. As shown in Fig. \ref{fig:topology}, there are two curves passing through the critical point. The one corresponding to a subsonic flow at $z_0$ is the one relevant for our problem. The other one corresponds to an accretion (infall) solution. The curves in the lower branch of solutions shown in Fig. \ref{fig:topology} correspond to flows that remain subsonic, the so called ``breezes". The upper branch corresponds to supersonic flow and is physically irrelevant. For both families of curves, there is a point where the numerator of equation (\ref{eq:Hcal-wind eq}) vanishes.  The other two branches are unphysical and for both there is a point where the denominator of equation (\ref{eq:Hcal-wind eq}) vanishes.

For given values of the magnetic field $B_0$, gas density $\rho_0$, gas pressure $P_{g0}$ and CR pressure $P_{c0}$ at $z=z_0$, we compute the velocity at the base of the wind $u_0$ that corresponds to the wind solution, namely the value of $u_0$ for which the flow starts as subsonic and then smoothly becomes supersonic at the critical point $z_c$ (see Fig. \ref{fig:topology}). Both the location of the sonic point and the launching velocity $u_{0}$ are outputs of the calculation. For each iteration, $P_{c0}$ and $\mathcal{D}(z)$ in equation (\ref{eq:Hcal-DD}) are computed based on the solution of the transport equation in the previous iteration. 

From the technical point of view, we start the search for the wind solution by bracketing the range of values of $u_{0}$ that may potentially correspond to a wind solution. In order to do so, we impose three conditions: 1) the flow starts as subsonic ($c_{*0}>u_0$); 2) u(z) is a growing function of $z$ at $z_{0}$, namely $\frac{du}{dz}\vert_{z_0} >0$; 3) the solution leads to a final wind velocity at infinity that is physical, namely $u_{f}^{2}>0$. 

This last point corresponds to imposing energy conservation between $z=z_{0}$ and infinity and requiring that at $z=\infty$ all fluxes vanish with the exception of the kinetic flux associated to the wind bulk motion. This condition selects the solutions that correspond to $u_{f}^{2}>0$. 

Once a closed interval for $u_0$, say $[u_{0,\rm min}, u_{0,\rm max}]$, has been determined, we numerically integrate the hydrodynamic equations starting at $z_0$ and for values of $u_0$ in this range, stepped by $\Delta u_0$. In this way we sample the topology of the solutions and we can identify the transition between the last breeze and the first unphysical solution by gradually increasing the value of $u_{0}$. This procedure can be repeated to narrow down the region where the actual wind solution is located. 


\begin{figure}
	\includegraphics[width=\columnwidth]{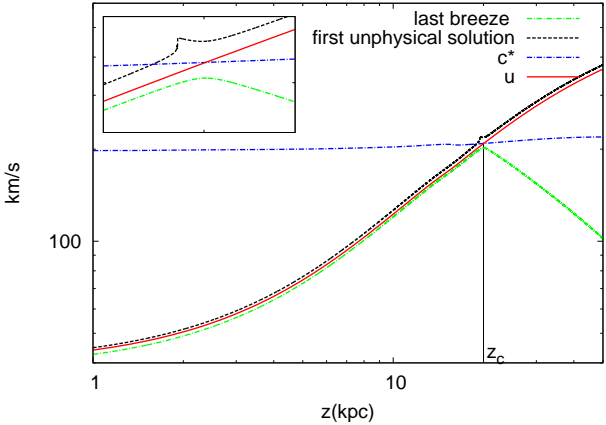}
    \caption{Schematic picture of the solution method for the hydrodynamic equations: the wind curve lies between a breeze, obtained by integrating with $u_0=u_{0B}$, and an unphysical solution, obtained by integrating with $u_0=u_{0U}$, where $u_{0B} < u_{0U}$. The location of the sonic point $z_c$ can be estimated as the location of the maximum of the breeze. The small box shows a zoomed region around the sonic point.}
    \label{fig:check-topology}
\end{figure}


The procedure of identification of the wind solution also returns the location of the critical point $z=z_{c}$. We can then compute both the flow speed and its derivative at the sonic point: 
\begin{align}
& u^2_c= \left. c^2_{*c}=\frac{d\Phi}{dz}\right/ \frac{1}{A} \frac{dA}{dz}, \label{eq:Hcal-uc} \\
& \left(\frac{du}{dz}\right)_{z_c}= \frac{u_c}{2}\left(\frac{1}{A}\frac{dA}{dz}\right)_{z_c} \\ \nonumber
& \qquad\qquad + \sqrt{\left(-\frac{u^2_c}{4}\left(\frac{1}{A}\frac{dA}{dz}\right)^2 + \frac{u^2_c}{2}\frac{1}{A}\frac{d^2A}{dz^2} -\frac{1}{2}\frac{d^2\Phi}{dz^2} \right)_{z_c}}. \label{eq:Hcal-dudzc}
\end{align}  
The last expression has been obtained by Taylor expanding equation (\ref{eq:Hcal-wind eq}) around the critical point. In addition, using the conservation relations (\ref{eq:Hcal-mass}), equations (\ref{eq:Hcal-B}) and (\ref{eq:energy cons}) and the fact that at the sonic point $u_c=c_{*}(z_c)$, we can compute the values of all physical quantities ($\rho$, $B$, $P_g$, $P_c$) at the sonic point. The wind solution is finally found by integrating the hydrodynamic equations starting at the sonic point toward $z=z_0$ and toward $z\to \infty$. 
The hydrodynamic part of our calculation has been checked versus the results of \cite{Breitschwerdt:1991A&A...245...79B} for the cases considered there and for the same set of parameters.


\section{Kinetic calculation}
\label{sec:Kin}

The stationary CR transport equation is as reported in equation (\ref{eq: CR transport equation flux tube}). Injection of CRs is assumed to take place only in an infinitely thin Galactic disc of radius $R_{d}$, so that the injection term is written as $Q(z,p)=Q_0(p)\delta(z)$, where
\begin{equation}\label{eq:injection}
Q_0(p)=\frac{\mathcal{N}_{SN}(p)\mathcal{R}_{SN}}{\pi R_d^2},
\end{equation}
and $\mathcal{N}_{SN}(p)$ is the spectrum contributed by an individual source occurring with a rate $\mathcal{R}_{SN}$. We have in mind supernova remnants (SNRs) as the sources of Galactic CRs, although the calculations presented here do not depend crucially on such an assumption. The spectrum of each SNR can be written as:
\begin{equation}
\mathcal{N}_{SN}(p)=\frac{\xi_{CR}E_{SN}}{I(\gamma)c(mc)^4}\left(\frac{p}{mc} \right)^{-\gamma},
\end{equation}
where $\xi_{CR}$ is the CR injection efficiency (typically $\sim 10\%$), $E_{SN}$ is the energy released by a supernova explosion ($\sim 10^{51}$erg), $\mathcal{R}_{SN}$ is the rate of SN explosions ($\sim 1/30 \, $yr$ ^{-1}$), and $I(\gamma)$ is a normalization factor chosen in such a way that 
\begin{equation}
\int_{0}^{\infty}\mathcal{N}_{SN}T(p) \, d^3p=\xi_{CR}E_{SN},
\end{equation}
where $T(p)$ is the kinetic energy of a particle of momentum $p$. 

Equation (\ref{eq: CR transport equation flux tube}) can be recast in a simpler form by noting that
\begin{equation}
UA\frac{\partial f}{\partial z}=\frac{\partial (AUf)}{\partial z}-f\frac{d(AU)}{dz},
\end{equation}
so that equation (\ref{eq: CR transport equation flux tube}) becomes
\begin{equation}
\frac{\partial}{\partial z} \left[ AD\frac{\partial f}{\partial z} -UAf\right] 
+\frac{d[AU]}{dz}\left[f+ \frac{p}{3}\frac{\partial f}{\partial p}\right] +AQ_0(p)\delta(z)=0. 
\end{equation}
In addition, 
\begin{equation}\label{eq: f vs dlnf}
 \frac{1}{3}\frac{1}{p^3}  \frac{\partial(fp^3) }{\partial \ln p} = \frac{f}{3}\frac{\partial\ln (fp^3)}{\partial\ln p} 
 = \frac{1}{3} \frac{\partial f}{\partial \ln p} + f 
\end{equation}
so that we can finally rewrite equation (\ref{eq: CR transport equation flux tube}) as
\begin{equation}\label{eq: CR transport def q}
\frac{\partial}{\partial z} \left[ AD\frac{\partial f}{\partial z} -UAf\right] 
-\frac{d[AU]}{dz}\frac{f}{3}q +AQ_0(p)\delta(z)=0,
\end{equation}
where we introduced the slope:
\begin{equation}\label{eq: def q}
q(z,p)=-\frac{\partial\ln (fp^3)}{\partial\ln p}.
\end{equation}
We look for an implicit solution of the transport equation (\ref{eq: CR transport def q}) that satisfies the boundary conditions $f(0,p)=f_0(p)$ at the disk and $f(p,H)=0$ at some outer boundary $H$. Note that here $H$ is set just for numerical purposes, and can be also located at spatial infinity. In other words the transport equation in the presence of a wind does not require the existence of a predefined physical halo of given size, in contrast with the standard calculations of CR transport, where such quantity plays a crucial role. 

The equation for $f_0(p)$ can be obtained by integrating equation (\ref{eq: CR transport def q}) around the disk, namely between $z=0^-$ and $z=0^+$, and using the fact that the problem is symmetric around the disk so that:
\begin{align*}
& U(0^+)= -U(0^-)=U_0 \\
& \left. \frac{dU}{dz}\right|_0=2U_0\delta(z)\\
&f(0^+)=f(0^-)=f_0(p)\\
&\left. \frac{dU}{dz}\right|_{0^+}= -\left. \frac{dU}{dz}\right|_{0^-}.
\end{align*}  
Integration around the disc leads to:
\begin{equation}\label{eq:boundary0}
2A_0D_0 \left. \frac{dU}{dz}\right|_{0^+} - 2 A_0U_0\frac{f_0}{3}\tilde{q}_0 + A_0Q_0(p)=0,
\end{equation}
where we used equations (\ref{eq: def q}) and (\ref{eq: f vs dlnf}) and we introduced
\begin{equation}\label{eq: def q0 tilde}
\tilde{q}_0(p)=-\frac{\partial\ln f_0(p)}{\partial\ln p}.
\end{equation}

Integrating equation (\ref{eq: CR transport def q}) between $z=0^+$ and a generic height $z$ one gets the equation:
\begin{align}
&\frac{\partial f}{\partial z}(z,p)= \frac{U(z)}{D(z,p)}f(z,p)\label{eq: dfdz}\\ \nonumber
&\qquad \qquad +\frac{\mathcal{G}(z,p)}{A(z)D(z,p)}-\frac{A_0 Q_0(p)}{2A(z)D(z,p)} -\frac{U_0A_0f_0(p)\overline{q}_0(p)}{A(z)D(z,p)},
\end{align}
where 
\begin{align}
& \mathcal{G}(z,p)=\int_0^z dz\; \frac{d(AU)}{dz}\frac{f}{3}q \label{eq: G storta}\\
& \overline{q}_0(p)=1-\frac{\tilde{q}_0}{3}\label{eq: q bar0}.
\end{align}
Integration of this equation between $z$ and $H$ with the boundary condition $f(H,p)=0$ leads to the following implicit solution for $f(z,p)$:
\begin{equation}\label{eq:f(z,p) solution}
f(z,p)={\int_z}^H \frac{dz^{\prime}}{A(z^{\prime})D(z^{\prime},p)} G(z^{\prime},p)
\exp^{-{\int_z}^{z^{\prime}}dz^{\prime\prime}\frac{U(z^{\prime\prime})}{D(z^{\prime\prime}, p)}},
\end{equation}
where
\begin{equation}\label{eq:G(z,p) solution}
G(z,p)=\frac{A_0 Q_0(p)}{2} + A_0U_0f_0(p)\bar{q}_0 - \mathcal{G}(z,p).
\end{equation}

This rewriting of the transport equations hides the non-linearity of the problem in the function $G(z,p)$, which depends on $f(z,p)$ and on the CR diffusion coefficient $D(z,p)$ (see equation \ref{eq:D self-gen}). The solution of equation (\ref{eq:f(z,p) solution}) is computed with an iterative procedure.

The procedure reaches convergence when, for a given iteration j, $f^j(z,p)$ and $f^{j-1}(z,p)$ are close to each other within a desired precision. Note that the advection velocity $U(z)$ is computed from the hydrodynamic equations and is fixed while iterating upon the distribution function $f^{j}$.


\section{Results}
\label{sec:results}

In this section we illustrate some selected cases of CR induced winds, aimed at addressing different issues that arise in this type of problem. The first case is what we will refer to as our {\it reference case}, namely a case that illustrates the most basic characteristics of a CR induced wind, with a minimal number of physical parameters introduced. The wind is launched very close to the disc of the Galaxy, and we consider the specific situation expected at the solar radius, namely at the Galactocentric distance $R=R_{\odot}=8.5$ kpc, where the Sun is located. This information is crucial in that it determines the gravitational potential that the wind has to fight against. The second model considered below is that in which a wind is launched at some distance from the Galactic disc, while particle transport in the near-disc region is assigned. This latter situation is physically motivated by the fact that ion-neutral damping is expected to damp self-generated Alfvenic turbulence within $\sim 0.5-1$ kpc from the disc, because of the presence of a substantial amount of neutral hydrogen. We will show that the consequences of this setup for the spectrum of CRs observed at the Earth are very prominent. 

\subsection{Reference case}
\label{sec:reference}

\begin{figure}
	\includegraphics[width=\columnwidth]{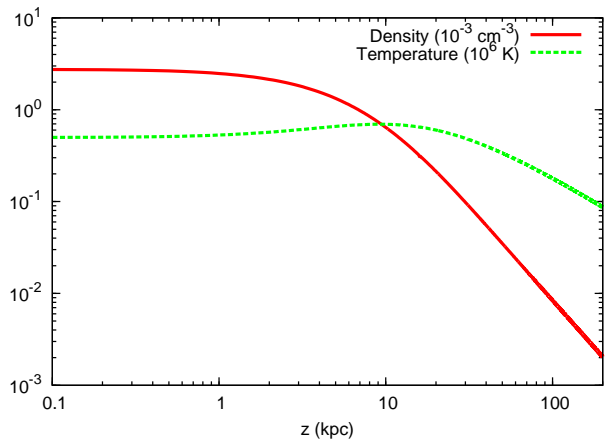}
    \caption{Density (red solid line) in units of $10^{-3}~\rm cm^{-3}$ and temperature (greed dashed line) in units of $10^{6}$ K for the wind solution in the reference case.}
    \label{fig:density}
\end{figure}

\begin{figure}
	\includegraphics[width=\columnwidth]{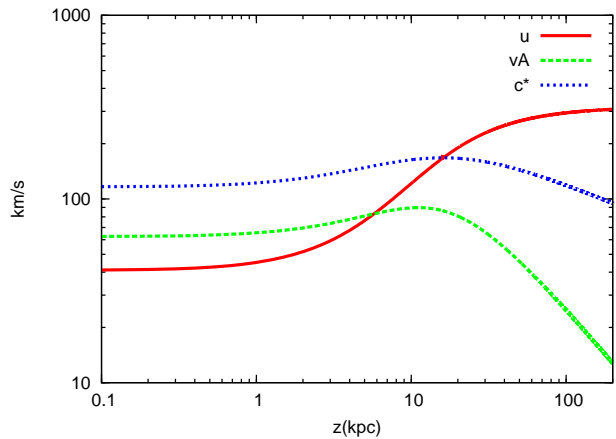}
    \caption{Wind velocity (red solid line), Alfv\'en velocity (green dashed line) and sound speed (blue dotted line) as a function of the height $z$ above the disc, for the reference case.}
    \label{fig:velocity}
\end{figure}

Our reference case corresponds to launching a CR-induced wind at a Galactocentric distance $R=R_{\odot}=8.5$ kpc. The base of the wind is assumed to be at the edge of the Galactic disc, $z_{0}=100$ pc, where we assumed that the ionized gas has a density $n_{0}=0.003~\rm cm^{-3}$ and the magnetic field is $B_{0}=1.5\mu G$ (to be interpreted as the component of the field along the $z$ direction). The CR pressure at $z_{0}$ is taken to equal the observed CR pressure, $P_{c0}=6\times 10^{-13}~\rm erg/cm^{3}$. We solve simultaneously the hydrodynamical equations for the wind and the transport equation for CRs, with self-generated diffusion and advection taken into account. In order to get the desired CR pressure at the Sun's location (see above) we are bound to take 
$$
\frac{\xi_{CR}}{0.1} \frac{\mathcal{R}_{SN}}{1/30 \rm \, yr^{-1}} \approx 1.8,
$$
for an injection spectrum with slope $\gamma=4.3$. The density of plasma in the wind and the temperature of the wind are shown in Fig. \ref{fig:density}. The temperature of the wind is maintained by the continuous damping of wave energy into thermal energy of the gas. In Fig. \ref{fig:velocity} we show the wind velocity $u(z)$ (green dashed line), the Alfv\'en speed $v_{A}(z)$ (blue dotted line) and the sound speed $c_{*}(z)$ as functions of the distance from the Galactic disc. The wind is launched with a speed of $41$ km s$^{-1}$ and asymptotically reaches a speed of $353$ km s$^{-1}$, while it becomes supersonic at $\sim 15$ kpc. 


For the sake of future discussion it is important to notice here that the CR advection velocity at the base of the wind is dominated by the Alfv\'en speed and that the latter is non zero at the base of the wind (because of a finite density and magnetic field). 

\begin{figure}
	\includegraphics[width=\columnwidth]{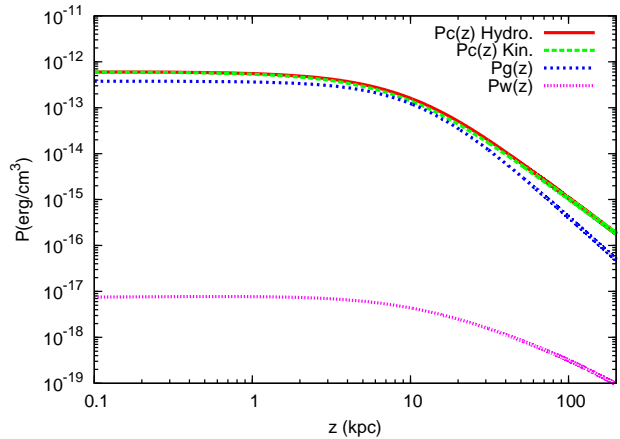}
    \caption{CR pressure (red solid line) and gas pressure (blue dotted line) in the reference case. The green dashed line shows the CR pressure as obtained from the solution of the transport equation. The same calculation also returns the wave pressure $P_{w}(z)$ (pink dotted line) that is assumed to vanish in the hydrodynamical equation.}
    \label{fig:Pc}
\end{figure}

The pressure of CRs as a function of the distance $z$ from the disc is shown in Fig. \ref{fig:Pc}, together with the gas pressure and the CR pressure as derived from the kinetic calculation: the fact that the latter is basically overlapped to $P_{c}(z)$ as derived from hydrodynamics shows that the system of equations (hydro plus kinetic) reached convergence (with accuracy of $\sim 0.1\%$ at heights $z$ close to the base of the wind and $\sim 10-20\%$ at $z$ close to the outer edge of the box used for numerical computation; here all quantities, with the exception of the wind velocity, are bound to vanish).\\
\\
The gas temperature obtained for $z \lesssim 10$ kpc (see Fig.~\ref{fig:density}) suggests that cooling may be important. At such temperature the gas is cooled by the emission of forbidden lines and soft X-rays \citep{Dalgarno1972} but, on the other hand, it is also heated by SN explosions  through the injection of hot gas and magnetic turbulence that will eventually decay into thermal energy. Recently it was suggested that also Coulomb losses of CRs themselves might be a substantial source of heating \citep{Walker2016}. 
The relevance of cooling was already recognized by \cite{Breitschwerdt:1991A&A...245...79B} which, nevertheless, assumed that energy balance exist between heating and cooling processes.  Which process dominates the heating is at the moment unclear but
observations reveals that the temperature in the halo reaches $\sim 10^5-10^6$ K \cite[see, e.g.,][]{Miller:2013ApJ...770..118M}, hence supporting the idea that either the cooling is negligible or it is balanced by heating processes.
In this paper we decided to neglect the role of cooling, deferring a more complete study in a future work.

\begin{figure}
	\includegraphics[width=\columnwidth]{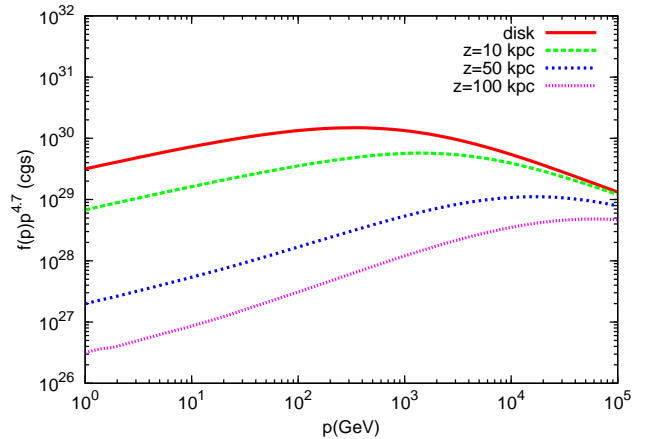}
    \caption{CR spectrum, $f(p)\times p^{4.7}$, in the reference case, at different locations in the wind: base of the wind (red solid line), $z=10$ kpc (green dashed line), $z=50$ kpc (blue short-dashed line) and $z=100$ kpc (pink dotted line).}
    \label{fig:fp}
\end{figure}

The spectrum of CRs as a function of particle momentum is reported in Fig. \ref{fig:fp} for several distances from the Galactic plane. The most striking feature of the spectrum at the disc is the pronounced spectral hardness: the spectrum that should be measured at the Earth location at momenta below $\sim 1$ TeV/c is roughly $\propto p^{-4.4}$, only slightly steeper than the injection spectrum ($Q(p)\propto p^{-4.3}$). This finding reflects the fact that the Alfv\'en speed at the base of the wind is very large and dominates CR advection up to high energies. One could argue that a larger value of the density at the base of the wind would make the Alfv\'en speed smaller, but in that case two problems appear: 1) it may become harder if not impossible to launch the wind because of excessive baryonic load. In other words, for a given CR pressure at the base, there may be cases in which the wind is not launched. 2) When the wind is in fact launched, its initial velocity may easily be super-Alfvenic, so that again advection dominates up to relatively high energies.  

This result on the spectrum of Galactic CRs appears to be at odds with previous results by \cite{Ptuskin:1997A&A...321..434P}, where a toy model for the velocity scaling with $z$ provided different results, of apparently easy interpretation. Since the argument of \cite{Ptuskin:1997A&A...321..434P} is rather simple, we report it here and we explain why this simple approach does not apply to realistic cases of CR-driven winds. The basic assumption of \cite{Ptuskin:1997A&A...321..434P} is that the advection velocity (dominated by the Alfv\'en speed) scales approximately linear with $z$, $v_{A}\sim \eta z$. Now, it is easy to imagine that while the advection velocity increases with $z$, it reaches a critical distance, $s_{*}$, for which advection dominates upon diffusion. This happens when
\begin{equation}
\frac{s_{*}^{2}}{D(p)}\approx \frac{s_{*}}{v_{A}(s_{*})} \, \Rightarrow \, s_{*}(p) \propto D(p)^{1/2},
\end{equation}
where we used the assumption of linear relation $v_{A}\sim \eta z$. Now, when diffusion dominates, namely when $z\lesssim s_{*}(p)$, one can neglect the advection terms and make the approximate statement (as in the standard diffusion model), that $D(p)\frac{\partial f}{\partial z}|_{z=0}\approx -Q_{0}(p)/2 \propto p^{-\gamma}$. Now, using equations (\ref{eq:wave spectrum}) and (\ref{eq:D self-gen}) one can easily show that $D(p)\propto p^{2\gamma-7}$ (for a $p^{-4}$ injection one gets a linear scaling of the diffusion coefficient with momentum). The quantity $s_{*}(p)$ plays the role of the size of the diffusion volume and one can show that, similar to a leaky box-like model, the equilibrium spectrum in the disc is 
\begin{equation}
f(p)\sim \frac{Q_{0}(p)}{s_{*}(p)}\frac{s_{*}^{2}}{D(p)} \sim Q_{0}(p)\, D(p)^{-1/2} \sim p^{-2\gamma + 7/2}.
\end{equation}
For injection $Q_{0}(p)\sim p^{-4.3}$ one infers an equilibrium spectrum $f_{0}(p)\sim p^{-5.1}$ (notice the contrast with the standard diffusion model within a given halo of size $H$, that predicts $f_{0}(p)\sim Q_{0}(p)H/D(p)$). 

\begin{figure}
	\includegraphics[width=\columnwidth]{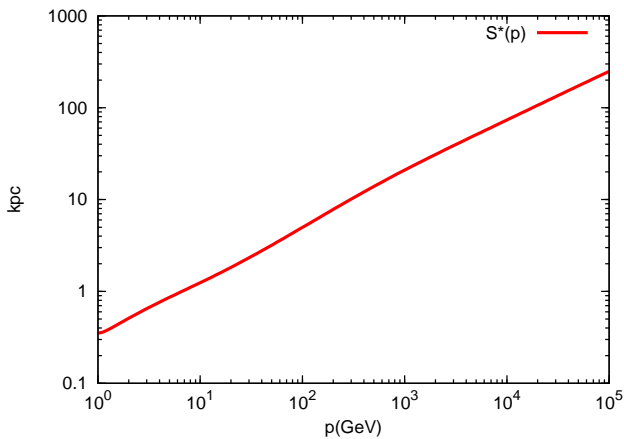}
    \caption{Effective boundary, $s_{*}(p)$, between the diffusion dominated and the advection dominated region of the wind in the reference case.}
    \label{fig:Sstar}
\end{figure}

\begin{figure}
	\includegraphics[width=\columnwidth]{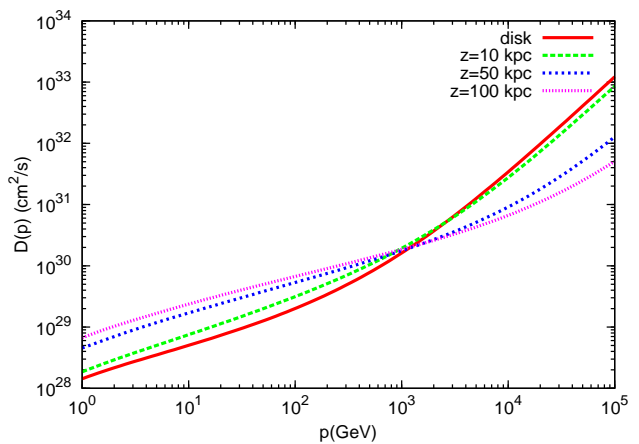}
    \caption{Self-generated diffusion coefficient, $D(p)$, in the reference case at different locations in the wind: base of the wind (red solid line), $z=10$ kpc (green dashed line), $z=50$ kpc (blue short-dashed line) and $z=100$ kpc (pink dotted line). The transition from the cylindrical to the spherical geometry of the wind flow is clearly visible in the momentum dependence of $D(p)$.}
    \label{fig:Dp}
\end{figure}

The problem with this argument, put forward by \cite{Ptuskin:1997A&A...321..434P}, is that it is strictly valid only when the advection velocity vanishes while approaching the base of the wind. One can see from equation (\ref{eq:boundary0}) that the assumption that diffusion is the dominant process at $z\to 0$ holds true only if $U_{0}=0$. The solution of the combined hydrodynamical equations and transport equation of CRs as presented above clearly shows that this is not the case. At low energies the slope of the CR spectrum at the base of the disc, as shown in Fig. \ref{fig:fp}, is $\sim 4.3$ (see also Fig. \ref{fig:f0}, lower panel), basically the same as the injection spectrum. The same point can be also made by plotting $s_{*}(p)$ (Fig. \ref{fig:Sstar}) and the diffusion coefficient $D(p)$ (Fig. \ref{fig:Dp}): the simple scaling $s_{*}(p)\propto D(p)^{1/2}$ can be easily seen to be not satisfied by the actual solution of the problem. Notice that $s_{*}(p)$ becomes larger than $Z_{s}$, the location where there is a transition to a spherical-like flow, at $p\sim 1$ TeV/c. At energies much larger than this one can assume that the advection velocity tends to a constant, $u_{f}$. The equilibrium spectrum observed at the Earth can then be written as $f_{0}(p)\sim Q_{0}(p)\pi R_{d}^{2}/(u_{f} s_{*}^{2})$, while $s_{*}\propto D(p)$, so that the equilibrium spectrum has a slope $-3\gamma+7$. This effect corresponds to a steepening of the equilibrium spectrum at the transition energy, that for the values of the parameters used in Fig. \ref{fig:fp}, corresponds to about $\sim 1$ TeV. 

Even qualitatively one can see that the CR spectrum in Fig. \ref{fig:fp} is quite different from the one actually observed at the Earth: it is harder than observed at low energies, and it is softer than observed at high energies, even though the pressure carried by these CRs is as observed. 

This example clearly shows that it is possible to construct solutions of the hydrodynamical equations that correspond to CR driven winds, with pressures at the base of the wind that are compatible with observations and yet leading to CR spectra that are not compatible with the CR spectra observed at the Earth. In particular, the basic wind model discussed in this section does not lead to any hardening of the spectrum at high energy, hence it is not possible to fit spectral hardening such as the ones observed by PAMELA \cite[]{pamhard} and AMS-02 \cite[]{amshard}. In \S \ref{sec:kolmo} below we discuss a situation in which this conclusion may not apply. 

\subsection{The importance of the near-disc region}
\label{sec:kolmo}

As already pointed out in the original work on CR driven winds  \cite[]{Breitschwerdt:1991A&A...245...79B} the region close to the disc may be plagued by severe ion-neutral damping which suppresses the generation of Alfv\'en waves \cite[]{ind}. Since waves provide the coupling between CRs and the ionized plasma, their severe damping leads to quenching of the wind. In fact, ion-neutral damping was recognized as a hindering factor for diffusion even in the absence of winds \cite[]{Skilling:1971p2173,Holmes:1975p621}: in these pioneering papers, the near disc region was assumed to be wave-free and the propagation of CRs in that region was taken to be ballistic. Diffusion in the outer halo, where the density of neutral hydrogen is expected to drop and the role of ion-neutral damping to become negligible, was considered as the actual diffusion region. One could however speculate that some type of turbulence may be maintained in the near-disc region, perhaps due to SN explosions themselves, though the waves may be considered to be isotropic, so that the effective Alfv\'en speed vanishes. 

In this section we discuss a scenario for the wind launching constructed in the following manner: the wind is assumed to be launched at a distance $z_{0}=1$ kpc from the disc and the near-disc region ($|z|<z_{0}$) is assumed to be characterised by a given diffusion coefficient, in the following form: 
\begin{equation}
D(p) = 3\times 10^{28} \left(\frac{p}{3 m_{p} c}\right)^{1/3} ~ \rm cm^{2} \, s^{-1},
\end{equation}
and by an Alfv\'en velocity $v_{A}=0$. At $z\geq z_{0}$, namely in the wind region, the diffusion coefficient is calculated as due to self-generation through streaming instability, saturated by NLLD, as in the reference case (\S \ref{sec:reference}). 

It is important to emphasise that the near-disc region is crucial to establish a connection between the sources and the wind region. From the mathematical point of view this is evident from equation (\ref{eq:boundary0}), where the solution of the transport equation in the disc, $f_{0}(p)$, is related to the injection rate through the value of the advection velocity at $z=0$. When such advection velocity is non zero, there is always a range of values of the particle momentum (at low momenta) where advection is more important than diffusion and the equilibrium spectrum turns out to have roughly the same slope as the injection spectrum. In the case that we consider in this section, we are assuming that the near-disc region is characterised by a vanishing advection velocity and finite assigned diffusion coefficient. From the technical point of view, the computation is the same as described above. The only minor difference is that, in order to avoid discontinuities in the advection velocity, that is zero in the near disc region but is not zero at $z=z_{0}$, we assume that both the wind velocity $u(z)$ and the Alfv\'en speed $v_{A}(z)$ have a low $z$ cutoff at $z<z_{0}$, so that both velocities drop to zero in a continuous manner in the near disc region. We checked that the details of such assumption do not have serious implications for the solution of the problem, provided that the velocity drops to zero fast enough inside the near-disc region. It is also worth stressing that, contrary to the reference case illustrated in \S \ref{sec:reference}, the CR pressure at the base of the wind ($z=z_{0}$) does not correspond to the CR pressure measured at the Earth ($z=0$). The criterion for convergence is still that the pressure at $z=0$ equals the observed CR pressure at the Earth location (clearly this would be different if we were interested at a different Galactocentric distance). 

\begin{figure}
	\includegraphics[width=\columnwidth]{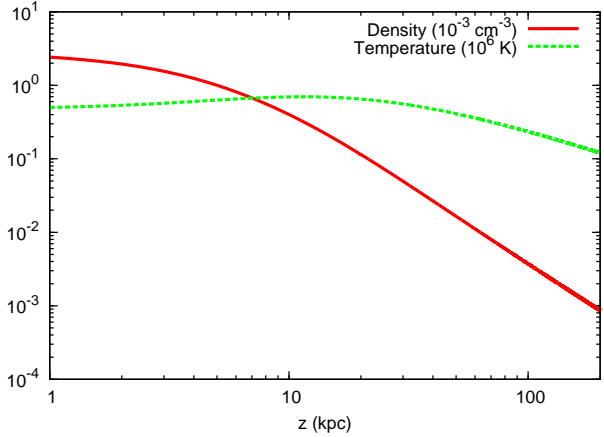}
    \caption{Density (red solid line) in units of $10^{-3}~\rm cm^{-3}$ and temperature (greed dashed line) in units of $10^{6}$ K limited to the region where the wind is launched, $z\geq z_{0}=1$ kpc.}
    \label{fig:densityK}
\end{figure}
\begin{figure}
	\includegraphics[width=\columnwidth]{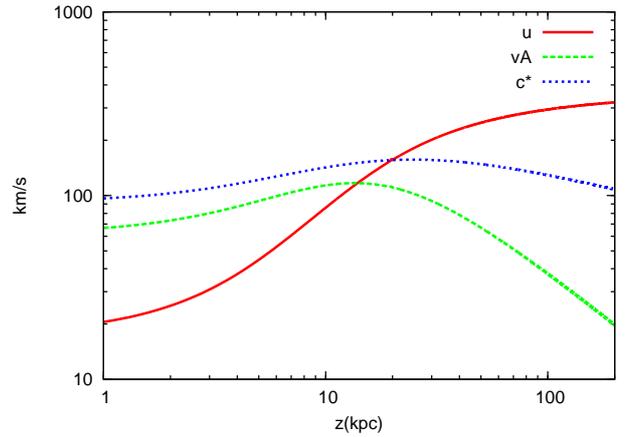}
    \caption{Wind velocity (red solid line), Alfv\'en velocity (green dashed line) and sound speed (blue dotted line) as a function of the height $z$ above the disc, limited to the region where the wind is launched, $z\geq z_{0}=1$ kpc.}
    \label{fig:velocityK}
\end{figure}
In order to recover the observed CR pressure in the Galactic disc at the Sun's location, in the model discussed here we need to require that:
$$
\frac{\xi_{CR}}{0.1} \frac{\mathcal{R}_{SN}}{1/30 \, \rm yr^{-1}} \approx 0.8.
$$

\begin{figure}
	\includegraphics[width=\columnwidth]{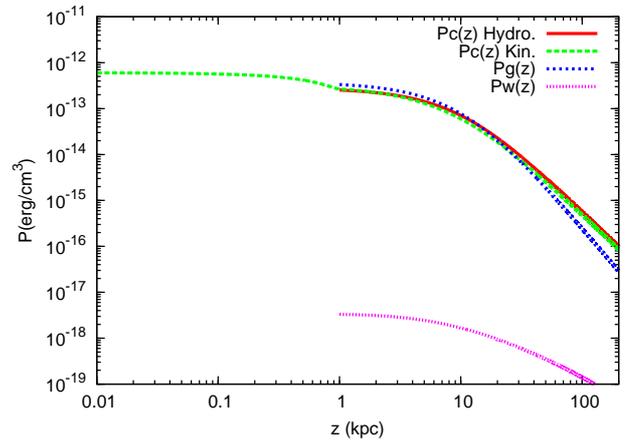}
    \caption{CR Pressure and gas pressure in the case of Kolmogorov turbulence in the near-disc region. The green dashed line shows the CR pressure as obtained from the solution of the transport equation. The same calculation also returns the wave pressure $P_{w}(z)$ that is assumed to vanish in the hydrodynamical equation.}
    \label{fig:PcK}
\end{figure}

The density and temperature of the wind, limited to the region $z\geq z_{0}=1$ kpc where the wind can be launched are plotted in Fig. \ref{fig:densityK}. The corresponding wind velocity, Alfv\'en speed and sound speed in the wind region are plotted in Fig. \ref{fig:velocityK}: the advection velocity is again dominated by the Alfv\'en velocity at the base of the wind. The wind becomes supersonic at $z_{c} \sim 20$ kpc, and eventually reaches $u_{f}\sim 400$ km/s at large distances from the disc. The pressures of the gas and CRs are shown in Fig. \ref{fig:PcK}, where the gas pressure is limited to the wind region while the CR pressure extends to the whole diffusion region. We also plotted there the CR pressure that is returned by the kinetic calculation, to show that with good accuracy it coincides with the one derived from the hydrodynamical calculation. The wave pressure as derived from the kinetic calculation is also shown: the plateau at $z<z_{0}$ has been estimated from the assigned diffusion coefficient. The fact that $P_{w}(z)$ is always much smaller than all other terms, justifies the assumption of instantaneous damping ($P_{w}=0$) in the hydrodynamical computation. 

\begin{figure}
	\includegraphics[width=\columnwidth]{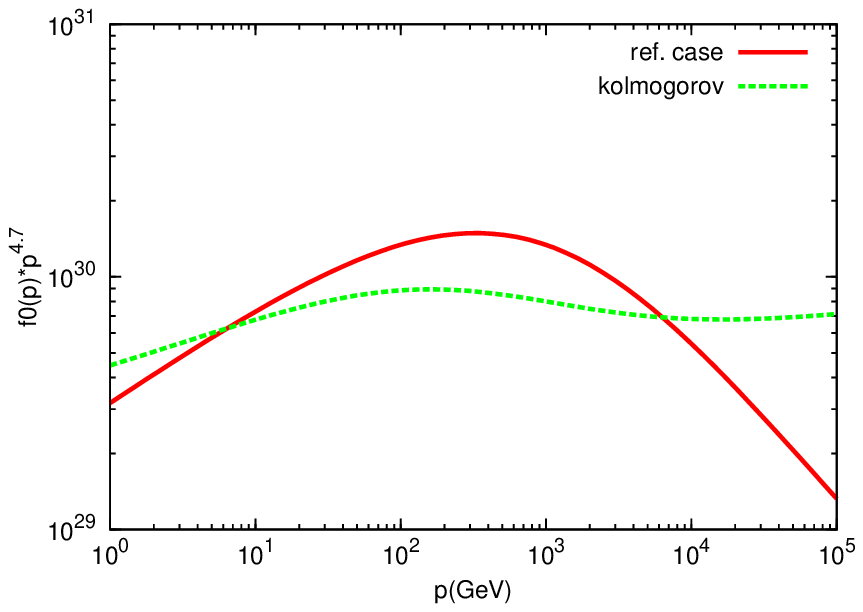}
	\includegraphics[width=\columnwidth]{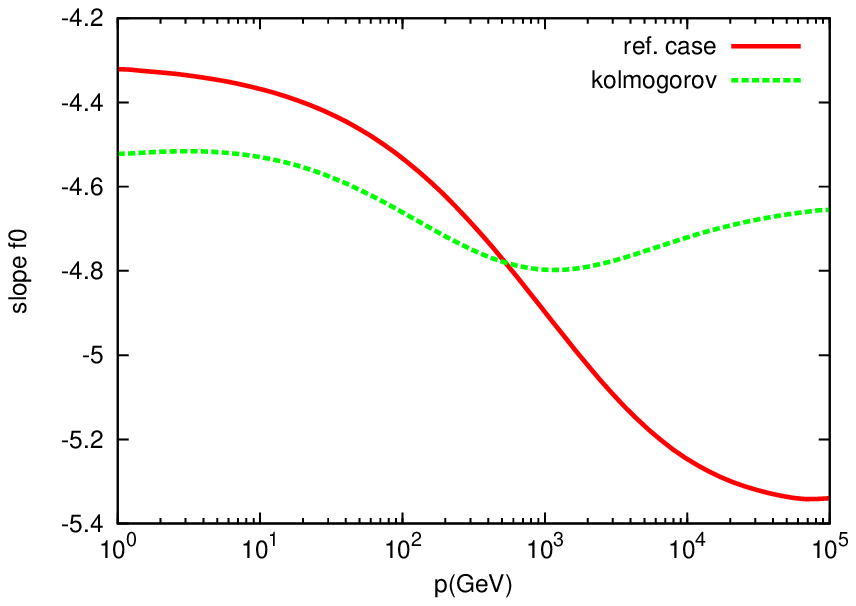}
    \caption{Upper panel: spectrum of CRs in the disc of the Galaxy at the position of the Sun for the reference case (red solid line) and for the case in which the wind is launched at $z_{0}=1$ kpc from the disc and there is a near-disc region where the diffusion coefficient is assigned and advection is absent (green dashed line). The lower panel shows the corresponding slopes.}
    \label{fig:f0}
\end{figure}

Using the approach described in \S \ref{sec:method}, \S \ref{sec:Hydro} and \S \ref{sec:Kin}, we also calculated the spectrum of Galactic CRs: in the upper panel of Fig. \ref{fig:f0} we show the CR distribution in the Galactic disc for the reference case compared to the case where the wind is launched at a finite distance $z_{0}$ from the disc. One can immediately see the dramatic difference that the near-disc region makes in terms of CR spectrum observed at the Earth (or, for that matter, anywhere). In the second case there are at least two spectral breaks that can be identified and explained in a reasonably easy way: the low energy part of the spectrum is affected by advection (below $\sim 10$ GeV) as well as from self-generation of waves at $z\geq z_{0}$ (at $10 \, \rm GeV\lesssim E\lesssim 1000$ GeV). At energies higher than $\sim 1$ TeV the spectrum hardens. All these changes of slope are illustrated more clearly in the lower panel of Fig. \ref{fig:f0}, where we show the slope of the CR distribution in the disc for the reference case (red solid line) and for the case with a wind launched at a finite distance $z_{0}$ from the disc (green dashed line). 

The hard spectrum (slope $\sim 4.5$ below $\sim 10$ GeV reflects advection and is similar to the effect already seen in non-linear models of galactic CR propagation \cite[see for instance]{Aloisio:2015p3650}. On the other hand, the spectrum has a continuous steepening at higher energies up to $\lesssim 1$ TeV: this effects is also similar to the one claimed by \cite{Aloisio:2015p3650} and reflects the very rapid energy dependence of the self-generated diffusion coefficient. Finally the slope of the spectrum becomes $\sim 4.63$ at $E\gtrsim 1$ TeV, because of the assumption of Kolmogorov turbulence in the near disc region ($z\le z_{0}$). This hardening is qualitatively similar to the one found by \cite{tomassetti} in a simple two zone model of the Galactic halo: the low energy part is sensitive to the far zone, while the high energy part is sensitive to the near zone, where the assumption of Kolmogorov turbulence leads to an expected slope $\rm 4.3 (injection) + 1/3 (Kolmogorov)\simeq 4.63$. The important point we wish to make here is that the diffusion coefficient that determines the low energy behaviour of the bulk of CRs is not assumed here, but rather derived from the condition of self-generation of waves and advection with the CR induced Galactic wind, as calculated above.

Although here we did not make any attempt to fit our curves to real data, it is interesting to notice that qualitatively the scenario presented in this section (contrary to the reference model in \S \ref{sec:reference}) has the same feature as the spectrum of CR protons as recently measured by PAMELA \cite[]{pamhard} and AMS-02 \cite[]{amshard}.

\section{Conclusions}
\label{sec:discuss}

We presented a semi-analytical calculation of the structure of a CR driven Galactic wind that returns the hydrodynamical structure of the wind (velocity and pressure of the plasma), and at the same time the spectrum of CRs at any location in the disc or in the wind region. The transport of CRs is described as advective and diffusive: advection occurs at a local speed that is the sum of the Alfv\'en speed and the wind speed at that point. Diffusion is due to scattering off self-generated turbulence, as generated by the same CRs due to streaming instability. The non-linear nature of the problem is manyfold: the wind is driven by the CR gradient in the $z$ direction, perpendicular to the Galactic disc, that results from the gradual escape of CRs from the Galaxy. The force $-\nabla P_{CR}$ acts in the direction opposite to gravity, so that, under certain conditions, it can lift ionized gas and launch a wind. On the other hand the amount of waves generated by CRs depends on their transport, which in turn depend on the wind speed. The diffusion coefficient is calculated in a self-consistent manner from the local spectrum of waves excited by streaming instability and damped through NLLD. 

Qualitatively, we confirm earlier findings that CRs can drive winds with asymptotic velocities of order several hundred km s$^{-1}$. Contrary to previous calculations, we also derived the spectrum of CRs as observed at the Earth (or any other location): for most of the parameter space for which a CR driven wind is launched, the spectrum is in disagreement with observations: the main reason for such a finding is that the CR advection with the wind is very strong and leads to hard spectra that are unlike the one observed at the Earth. However, this conclusion is found to depend in a critical way on the conditions in the near-disc region ($z\lesssim 1$ kpc). The conditions in such region affect the high energy behavior of the spectrum, and change the launching speed of the wind at $z\sim z_{0}\sim 1$ kpc, so that the low energy region of the spectrum is also changed. For certain assumptions on the diffusion coefficient in the near-disc region, the predicted CR spectrum may resemble the observed one: as an instance we showed one case in which the spectrum is relatively hard at $E\lesssim 10$ GeV  \cite[as previously found by][]{Aloisio:2015p3650}, steeper for $10\lesssim E\lesssim 1000$ GeV, and show a hardening at higher energies as recently claimed by \cite{pamhard,amshard}. From the physical point of view, we expect that the region at $z\lesssim 1$ kpc is reach in neutral gas, which may damp Alfv\'en waves through ion-neutral damping. This implies that the coupling between CRs and gas may be exceedingly weak in this region so as to inhibit the launch of the wind. In the near-disc region, diffusion, if any is present, should be guaranteed by some source other than CR self-generation. 

The calculation presented here is very general and can be applied to a variety of conditions. This will allow us to explore the conditions in which a CR driven wind could develop with a CR spectrum at the Earth resembling the observed one. This type of calculations is also useful to develop possible diagnostics of the existence of CR-driven winds in our own Galaxy as well as in others. 

\section*{Acknowledgements} The authors are glad to acknowledge a continuous ongoing collaboration with the rest of the Arcetri high energy Group. This work was partially funded through Grant PRIN-INAF 2012.



\bibliographystyle{mnras}
\bibliography{biblio} 



\appendix
\section{Derivation of the hydrodynamic equations}
\label{sec:appendix Hydro}
In this section we derive the hydrodynamic equations used throughout this paper. They describe the macroscopic behaviour of the thermal plasma and CRs and their mutual interactions. As we already pointed out in Sec. \ref{sec:method}, the wave component is not present in these equations because we are assuming that all Alfv\'{e}n  waves (generated through CR streaming instability) are locally dumped on short time scales, so that the energy fed by CRs into waves just results in the heating of the gas. Here we are also neglecting any external loss of mass, momentum and energy. A general form of those equations, with and without damping of waves, is reported in \cite{Breitschwerdt:1991A&A...245...79B}.

\subsection{Gas mass conservation and equation of motion}
In absence of any mass and momentum external loss, the time dependent gas mass conservation and equation of motion read
\begin{equation}\label{eq:Ht-mass}
\frac{D\rho}{Dt}=-\rho\vec{\nabla}\cdot\vec{u},
\end{equation}
\begin{equation}\label{eq:Ht-motion}
\rho\frac{D\vec{u}}{Dt}=-\vec{\nabla}(P_g +P_c) - \rho\vec{\nabla}\Phi,
\end{equation}
where $\rho$, $\vec{u}$, $P_g$, $P_c$ and $\Phi$ are the gas density, velocity and pressure, the CR pressure and the galactic gravitational potential respectively. $D/Dt = \partial/\partial t + \vec{u}\cdot\vec{\nabla}$ is the convective derivative.

\subsection{Gas internal energy and total energy}
The internal energy of the gas in a volume $V$ is given by
\begin{equation}
\epsilon_g=\frac{Pg V}{\gamma_g -1}
\end{equation} 
and its time evolution is given by
\begin{equation}
\frac{D\epsilon_g}{Dt}= -Pg\frac{DV}{Dt}- \vec{v}_A\cdot \vec{\nabla}P_c V
\end{equation} 
where on the right hand side we find the work done on the gas when $V$ changes and the energy input due to damping of the self generated Alfv\'{e}n waves. In fact, as it will be shown in \S~\ref{sec:waves}, the hydrodynamic counterpart of the CR streaming instability is given by the term $\vec{v}_A\cdot \vec{\nabla}P_c$, where $\vec{v}_A$ is the Alfv\'{e}n speed and $P_c$ the CR pressure. In terms of $\rho$, and using equation (\ref{eq:Ht-mass}), we get
\begin{equation}\label{eq:Ht-Pg}
\frac{D}{Dt}\left(\frac{Pg}{\gamma_g -1}\frac{1}{\rho} \right)= -\frac{P_g}{\rho}\vec{\nabla}\cdot\vec{u} - \frac{1}{\rho} \vec{v}_A\cdot \vec{\nabla}P_c.
\end{equation}
The gas total energy per unit volume is the sum of the kinetic and internal energy per unit volume 
\begin{equation}
\epsilon= \frac{1}{2}\rho u^2 +\frac{Pg}{\gamma_g -1}.
\end{equation} 
Making use of equations (\ref{eq:Ht-mass}), (\ref{eq:Ht-motion}) and (\ref{eq:Ht-Pg}), we get the equation for the gas total energy
\begin{align}
\frac{D}{Dt}\left(\frac{1}{2}\rho u^2 +\frac{Pg}{\gamma_g -1}\right) &= -\vec{\nabla}\cdot\vec{u}\left(\frac{1}{2}\rho u^2 
+ \frac{\gamma_g Pg}{\gamma_g -1}\right)
\label{eq:Ht-energy} \\ \nonumber
& -\vec{u}\cdot\vec{\nabla}(P_g + P_c) - \vec{v}_A\cdot \nabla P_c - \rho\vec{u}\cdot\vec{\nabla}\Phi.
\end{align}

\subsection{CR energy}
The time dependent CR transport equation reads
\begin{equation}
\frac{\partial f}{\partial t} +(\vec{u}+\vec{v}_A)\cdot\vec{\nabla}f= \vec{\nabla}\cdot\left[D\vec{\nabla}f\right]
	+ \left[ \vec{\nabla} \cdot (\vec{u}+\vec{v}_A) \right] \frac{p}{3} \frac{\partial f}{\partial p}
\end{equation}
where $f$ is the CR distribution function and $D$ the CR diffusion coefficient. We also introduce the CR energy density and pressure 
\begin{align}
&\epsilon_c = \int_0^\infty dp\; 4\pi p^2 T(p)f(z,p) ,\\
&P_c = \int_0^\infty dp\; \frac{4\pi}{3}  p^2 pvf(z,p), \label{eq:def-Pc}
\end{align}
where $T(p)=\sqrt{p^2c^2 + (mc^2)^2}-mc^2$ is the kinetic energy. Multiplying the transport equation by $4\pi\; p^2 T(p)$ and integrating in $p$ we get
\begin{align}
 & \frac{\partial \epsilon_c}{\partial t} +(\vec{u}+\vec{v}_A)\cdot\vec{\nabla}\epsilon_c    
  = \vec{\nabla} \cdot \left[\overline{D}\vec{\nabla}\epsilon_c\right] +    \nonumber \\
  & \hspace{2cm} + \vec{\nabla} \cdot(\vec{u}+\vec{v}_A) 
         \int_0^\infty \frac{4}{3}\pi dp p^3 T(p)\frac{\partial f}{\partial p}
\end{align} 
where we introduced the average diffusion coefficient
\begin{equation}\label{eq:average D}
\overline{D}(z)=\frac{\int_0^\infty dp\; 4\pi  p^2 T(p)D(z,p)\vec{\nabla}f}{\int_0^\infty dp\; 4\pi p^2 T(p)\vec{\nabla}f}.
\end{equation}
The last integral can be performed by parts, using the fact that $dT(p)/dp = v$. After rearranging terms and introducing the effective CR adiabatic index $\epsilon_c = P_c/(\gamma_c -1)$ we obtain: 
\begin{equation}\label{eq:Ht-CR}
\frac{\partial \epsilon_c}{\partial t} + \vec{\nabla}\cdot\left[(\vec{u}+\vec{v}_A)\frac{\gamma_c P_c}{\gamma_c -1} \right]=
\vec{\nabla}\cdot\left[\overline{D}\frac{\vec{\nabla} P_c}{\gamma_c -1} \right] + (\vec{u}+\vec{v}_A)\cdot \vec{\nabla}P_c.
\end{equation}

\subsection{Wave energy} \label{sec:waves}
The time evolution of the Alfv\'{e}n wave spectral energy density $W(\vec{z},p)$, taking into account wave advection by the thermal plasma and wave generation by CR streaming instability, is given in \cite{Jones:1993ApJ...413..619J}
\begin{equation}\label{eq:Kt-waves}
\frac{\partial W}{\partial t} + \vec{\nabla}\cdot\left[\left(\frac{3}{2}\vec{u}+\vec{v}_A\right)W\right] =\frac{1}{2}\vec{u}\cdot \vec{\nabla}W -\Gamma_{\rm CR}W
\end{equation}
where $\Gamma_{CR}$ (see also equation \ref{eq:GAMMA CR}) is the wave growth rate due to CR streaming instability ($k$ is the wave number and $B_0$ the average magnetic field)
\begin{equation}
\Gamma_{\rm CR}=\frac{16\pi^2}{3}\frac{\vec{v}_A}{kWB_0^2}\cdot\int_0^\infty dp p^4 v(p)\delta\left(p- \frac{qB_0}{kc}\right)\vec{\nabla}f.
\end{equation}
We also define the wave energy density and pressure
\begin{align}
&\epsilon_w = \int_{k_0}^\infty \frac{{B_0}^2}{4\pi} W(z,k)dk\\
& P_w=\frac{\epsilon_w}{2}.
\end{align}
The hydrodynamic version of the wave equation can be obtained by integrating in $\int_{k_0}^\infty \frac{{B_0}^2}{4\pi}dk$. The integration of the streaming instability term gives
\begin{align}
& \int_{k_0}^\infty \frac{{B_0}^2}{4\pi}\Gamma_{\rm CR}W(z,k)dk \\ \nonumber
& \qquad =\int_{0}^\infty \int_{k_0}^\infty \frac{4\pi}{3}
\frac{\vec{v}_A}{k}\cdot \vec{\nabla}f p^4 v(p) \, \delta\left(p- \frac{qB_0}{kc}\right)dpdk \\ \nonumber
& \qquad = \int_{0}^\infty \int_{k_0}^\infty \frac{4\pi}{3}
\vec{v}_A\cdot \vec{\nabla}f p^4 v(p)\frac{kc}{qB_0}  \, \delta\left(k- \frac{qB_0}{pc}\right)dpdk \\ \nonumber
& \qquad =\vec{v}_A\cdot \vec{\nabla}\int_{0}^\infty \frac{4\pi}{3} f v(p)p^3dp\\ \nonumber
& \qquad =\vec{v}_A\cdot \vec{\nabla}P_c
\end{align}
where we used the properties of the Dirac's delta and the definition of $P_c$ given in equation (\ref{eq:def-Pc}). Finally, the equation for the wave energy is
\begin{equation}
\frac{\partial \epsilon_w}{\partial t} + \vec{\nabla}\cdot\left[\left(3\vec{u}+2\vec{v}_A\right)P_w\right] =\vec{u}\cdot \vec{\nabla}P_w -\vec{v}_A\cdot \vec{\nabla}P_c.
\end{equation}
In our approach we assume that all generated waves are locally and immediately dumped, so that the hydrodynamic equation  for waves is not necessary. On the other hand, the term of wave generation by CR streaming instability, $\vec{v}_A\cdot \vec{\nabla}P_c$, enters the right hand side of the equation for the gas pressure (\ref{eq:Ht-Pg}) because the damped waves result in gas heating.

\section{Stationary hydrodynamic equations in the flux tube}
\label{sec:appendix_tube}
We are dealing with a steady one-dimensional flow, in which the flux tube geometry is preassigned as shown in Fig. \ref{fig:flux-tube}. The stationary regime is obtained by setting $\partial/\partial t =0$ in the hydrodynamic equations. The stationary hydrodynamic equations are listed in \S~\ref{sec:method}. As for the flux tube geometry, referring to Fig. \ref{fig:flux-tube}, if $z$ is the vertical coordinate and $A(z)$ the flux tube area, the divergence and gradient operators become
\begin{align}
&\vec{\nabla}S=\frac{dS}{dz} \hat{z}\label{eq:gradient in flux tube}\\
&\vec{\nabla}\cdot \vec{V}=\frac{1}{A(z)}\frac{d}{dz}
\left[ \vec{A}(z)\cdot \vec{V} \right]. \label{eq:divergence in flux tube}
\end{align}   
By applying this prescription for the gradient and divergence operators to the stationary equations we get 
\begin{align}
& \rho u A=\textnormal{const}, \label{eq:Hft-mass} \\
& AB= \textnormal{const},\label{eq:Hft-B}\\
& \rho u \frac{d}{dz}\left(\frac{u^2}{2} + \Phi\right) = -u\frac{d}{dz}\left(P_g + P_c\right), \label{eq:Hft-u}\\
& \frac{dP_g}{dz}=\gamma_g\frac{P_g}{\rho}\frac{d\rho}{dz}-(\gamma_g -1)\frac{v_A}{u}\frac{dP_c}{dz} \label{eq:Hft-Pg}\\
&\rho uA \frac{d}{dz}\left(\frac{u^2}{2} +\frac{\gamma_g}{\gamma_g -1} \frac{P_g}{\rho} + \Phi\right) = -A(u+v_A)\frac{dP_c}{dz} , \label{eq:Hft-energy}\\
&\frac{dP_c}{dz}=\frac{\gamma_c P_c}{\rho}\frac{2u+v_A}{2(u+v_A-\mathcal{D})}\frac{d\rho}{dz}. \label{eq:Hft-Pc}
\end{align}
In the derivation of equation (\ref{eq:Hft-Pc}) we defined 
\begin{equation}
\mathcal{D}=\frac{\frac{d}{dz}\left(A\overline{D}\frac{dP_c}{dz}\right)}{A\frac{dP_c}{dz}}
\end{equation}
and we used 
\begin{align}
&\frac{d(uA)}{dz} = - \frac{uA}{\rho}\frac{d\rho}{dz}\label{eq:dAU} \\
&\frac{d(v_AA)}{dz} = - \frac{v_AA}{2\rho}\frac{d\rho}{dz}.\label{eq:dvAA}
\end{align}

\subsection{First integrals}
It is possible to derive from the hydrodynamic equations two first integrals which come in handy when searching for the wind solution. The first one derives directly from the total energy conservation: summing equations (\ref{eq:H-energy}) and (\ref{eq:H-Pc}) and moving to the flux tube geometry, we get
\begin{equation}\label{eq:energy cons}
\frac{u^2}{2} +\frac{\gamma_g}{\gamma_g -1} \frac{P_g}{\rho} + \Phi +\frac{\gamma_c}{\gamma_c -1} \frac{P_c}{\rho}\frac{u+v_A}{u}
-\frac{\overline{D}\frac{dP_c}{dz}}{(\gamma_c-1)\rho u}=\textnormal{const}.
\end{equation}
The other first integral is obtained from equation (\ref{eq:Hft-Pc}) when $\overline{D}=0$, i.e when $\mathcal{D}=0$. Due to this limitation, this first integral is only used in the very first iteration of our calculation, where we assume $\overline{D}=0$.

In fact, using equations (\ref{eq:dAU}) and (\ref{eq:dvAA})
\begin{equation}
\frac{d}{dz}\left(A(u+v_A)\right)=-\frac{A(2u+v_A)}{2\rho}\frac{d\rho}{dz},
\end{equation}
thus $dP_c/dz$ in equation (\ref{eq:Hft-Pc}) becomes
\begin{equation}
\frac{dP_c}{dz} = - \frac{\gamma_c P_c}{A(u+v_A)} \, \frac{d}{dz} \left[ A(u+v_A) \right]
\end{equation}
from which we finally get
\begin{equation}\label{eq:Pc cons}
P_c\left[A(u+v_A)\right]^{\gamma_c}=\textnormal{const}.
\end{equation}

\subsection{The wind equation}
Starting from equation (\ref{eq:Hft-u}) and using equations (\ref{eq:Hft-Pg}) and (\ref{eq:Hft-Pc}) we get the so-called wind equation, i.e an equation for the flow speed,
\begin{equation}
\frac{du}{dz}=u\frac{c_{*}^2\frac{1}{A} \frac{dA}{dz} -\frac{d\Phi}{dz}}{u^2-c_{*}^2}
\end{equation}
where we defined the ``compound sound speed"
\begin{equation}
c_{*}^2=\gamma_g \frac{P_g}{\rho} + \gamma_c \frac{P_c}{\rho} \left[ 1-(\gamma_g-1)\frac{v_A}{u} \right]\frac{2u+v_A}{2(u+v_A -\mathcal{D})}.
\end{equation}

\bsp	
\label{lastpage}
\end{document}